\documentstyle[stwol]{article}

\def\NPB{{\em Nucl. Phys.} B}
\def\PLB{{\em Phys. Lett.}  B}

\def\be{\begin{equation}}
\def\ee{\end{equation}}
\def\bea{\begin{eqnarray}}
\def\eea{\end{eqnarray}}

\begin{document}
\title{
\flushleft {\rm RI-8-96, NYU-TH-96/10/01, hep-th/9610159}  
\\ \vskip .2in \center
ASPECTS OF DUALITIES}

\author{AMIT GIVEON}

\address{Theory Division, CERN, CH-1211, Geneva 23, Switzerland}

\author{MASSIMO PORRATI}

\address{Dept. of Physics, NYU, 4 Washington Pl., New York NY 10003, USA\\
         and\\ Rockefeller Univ., New York NY 10021-6399, USA}


\twocolumn[\maketitle\abstracts{In this talk,
some aspects of duality symmetries are presented.}]

In the last few years, duality symmetries in field theory and in string theory
have been studied extensively. Let us list some of them:
\begin{itemize}
\item
{\em Electric-Magnetic Dualities}: strong-weak coupling duality which relates
apparently different field theories under the interchange of
electric degrees of freedom with magnetic degrees of freedom (for a review, see
\cite{o,s})

\item
{\em T-Duality}: target space duality in string theory relates different string
backgrounds which are physically equivalent (for a review, see \cite{gpr}).

\item
{\em S-Duality}: strong-weak coupling duality in string theory (for a review,
see \cite{sen}).

\item
{\em U-Duality}: intertwines T-duality and S-duality \cite{ht}.

\item
{\em String-String Dualities}: relate different string theories to each other
(for a review, see \cite{p}).

\item
{\em World-Sheet $\leftrightarrow$ Target-Space Dualities}: relate a theory
on a world-sheet and target-space $(\Sigma,T)$ to a theory on a different
world-sheet and a different target-space $(\tilde{\Sigma},\tilde{T})$
\cite{gmr}.

\end{itemize}

The new results in this talk\footnote{
Talk presented by Amit Giveon at the ICHEP `96, Warsaw, July 25-31, 1996}
are based on work in ref. \cite{gp}.
We present duality symmetries in $4D$, Abelian gauge theories
which involve also the (Euclidean, compact) space-time $M^4$.
Such dualities -- rather mysterious from the $4D$ point of view -- are better
understood as geometrical symmetries of theories in higher dimensions,
compactified to $M^4$ on some internal space.

Explicitly, we find dualities which relate a pair $(M^4,\tau)$ to a different
pair $(\tilde{M}^4,\tilde{\tau})$, where $\tau$ is the complex
coupling-constant matrix of a $U(1)^r$ gauge theory on $M^4$, and
$\tilde{\tau}$ is the dual coupling-constant matrix of a $U(1)^{\tilde{r}}$
gauge theory on $\tilde{M}^4$.

Some of the dualities considered can be understood as the consequence of string
dualities, in the limit where gravity is decoupled. A simple example is the
string-string triality which relates the heterotic string compactified on
$T^4\times T^2$ to Type II strings on $K^3\times T^2$. In the low-energy
effective field theory, at the limit $M_{Planck}\to \infty$, this triality
becomes part of the electric-magnetic duality group of an $N=4$,
$U(1)^4$ supersymmetric Yang-Mills (YM) theory on ${\bf R}^{3,1}$.

In the rest of the talk, we continue by touring more examples which involve
also space-time:

\begin{enumerate}

\item
{\em $S\leftrightarrow U$ Duality in $N=4$ YM Theory and Self-Dual Superstring
in $6D$}: consider $SU(2)$, $N=4$ YM theory on $M^4=S^1_{\beta}\times
S^1_{R}\times T^2_{U}$, where $\beta$ is the inverse temperature, $R$ is the
radius of a circle, and $T^2_U$ is a 2-torus with complex structure $U$.
Let $S$
denote the complex gauge coupling. The partition function at large scalar VEVs
was computed in \cite{ggpz}. It turns out that not only is it invariant under
$SL(2,{\bf Z})$, $S$-duality and the geometrical $SL(2,{\bf Z})$
transformations acting on the complex structure, but also under the
transformation interchanging $S$ with $U$. This $S\leftrightarrow U$ duality is
a mysterious symmetry from the $4D$ gauge theory point of view. As explained in
\cite{gp}, we can understand the origin of this duality from the manifest
geometrical symmetry of a compactified self-dual superstring in $6D$.
Explicitly, by reducing a theory of a self-dual 2-form in $6D$ to $2D$ on
$T^2_U\times T^2_S$, we obtain the same partition function as above.
The $S\leftrightarrow U$ duality is a consequence of the manifest
$T^2_U\leftrightarrow T^2_S$ geometrical symmetry.
The technical details are given in \cite{gp}.

\item
{\em Compactification of a 2-Form Theory in $6D$ and Triality}:
by compactifying a theory of a self-dual 2-form in $6D$ on
$T^2_T\times T^2_U\times T^2_S$, we obtain a triality symmetry under the
permutations of $S,T$ and $U$. In a certain large volume limit, the
partition function is identical to the classical
part of the one-loop partition function of a $2D$ sigma-model with a $T^2$
target-space. In string theory, this triality -- observed sometime ago
\cite{dvv} -- is rather mysterious, because $T$ is the complex structure of
the world-sheet torus, while $U$ and $S$ are the complex structure and the
K\" ahler structure of the target-space torus, respectively. However, for
the 2-form theory on $T^2_T\times T^2_U\times T^2_S$, this triality is the
geometrical symmetry permuting the three 2-tori.

\item
{\em  Compactifications of a 2-Form Theory in $6D$ and More Dualities}:
a generalization of the previous example is to compactify a
self-dual 2-form  theory in $6D$ on $\Sigma_g\times T^2_S\times
\tilde{\Sigma}_{\tilde{g}}$, where $\Sigma_g$ and $\tilde{\Sigma}_{\tilde{g}}$
are (different) Riemann surfaces with genus $g$ and $\tilde{g}$, respectively.
The partition function is identical to the classical partition function of a
$2D$ sigma-model with world-sheet $\Sigma_g$ (alternatively,
$\tilde{\Sigma}_{\tilde{g}}$) and target-space $T^{2\tilde{g}}$ (alternatively,
$T^{2g}$) whose metric and antisymmetric background are parametrized by $S$ and
the period matrix of $\tilde{\Sigma}_{\tilde{g}}$ (alternatively, by
the period matrix of $\Sigma_g$). We find a world-sheet $\leftrightarrow$
target-space duality between the pairs
$$ \{\Sigma_g,T^{2\tilde{g}}(\tilde{\Sigma}_{\tilde{g}})\}
\leftrightarrow
\{\tilde{\Sigma}_{\tilde{g}},T^{2g}(\Sigma_g)\}. $$
This duality is a manifest consequence of the geometrical symmetry
interchanging $\Sigma_g\leftrightarrow \tilde{\Sigma}_{\tilde{g}}$ in the
compactified 2-form theory.
Similar world-sheet $\leftrightarrow$ target-space dualities were observed in
\cite{gmr}. These can be understood as the  $\Sigma_g\leftrightarrow
\tilde{\Sigma}_{\tilde{g}}$ geometrical symmetries of a compactified $6D$
2-form theory, which instead of self-duality obey other conditions \cite{gp}.

\item
{\em $(M^4,\tau)\leftrightarrow (\tilde{M}^4,\tilde{\tau})$ Duality and
Compactifications of a Self-Dual 4-Form Theory in $10D$}:
a reduction of a $10D$ self-dual 4-form theory to $4D$ on $M^4\times T^2_S$
gives a $4D$, $U(1)^{b_2}$ gauge theory, where $M^4$ is a 4-manifold with
$b_1(M^4)=0$; $b_1,b_2$ are the Betti numbers of $M^4$.
Upon further compactification on $M^4\times T^2_S\times \tilde{M}^4$ we find a
manifest symmetry under the interchange $M^4\leftrightarrow  \tilde{M}^4$.
This translates into a rather ``puzzling'' duality of the $4D$ gauge theory:
$$
\{M^4,\tau(\tilde{M}^4)\}\leftrightarrow \{\tilde{M}^4,\tilde{\tau}(M^4)\},
$$
where $\tau(\tilde{M}^4)$ is the gauge-coupling matrix given in terms of the
geometrical data of $\tilde{M}^4$, and $\tilde{\tau}(M^4)$ is the dual
gauge-coupling matrix given in terms of the geometrical data of $M^4$.
This duality relates a theory on space-time manifold $M^4$ and
coupling-constant matrix $\tau$ to a theory with a different 4-manifold and a
different gauge-coupling matrix.
This construction sometimes works for more than the partition function of an
Abelian gauge theory. For $M^4=K^3$ and $\tilde{M}^4=\tilde{K}^3$, this
construction can be embedded in an $N=4$ compactification of type IIB string on
$K^3\times T^2_S\times \tilde{K}^3$.

To summarize:

\begin{itemize}
\item
Duality groups of Abelian gauge theories on 4-manifolds and their reduction to
$2D$ were considered.

\item
The duality groups include elements that relate different space-times in
addition to relating different gauge-coupling matrices.

\item
We interpreted such dualities as geometrical symmetries of compactified
theories in higher dimensions.

\item
In particular, we considered compactifications of a (self-dual) 2-form in $6D$,
and of a self-dual 4-form in $10D$.

\item
Relations with a (conjectured) self-dual superstring in $6D$ and with Type IIB
superstring were mentioned.

\item
There are many more duality symmetries of the classical partition sum of free
$U(1)^r$ gauge theories on $M^4$ which were not discussed here.

\item
To recover all symmetries of $4D$ gauge theories, and their possible
geometrical origin from higher dimensions, is a problem which may shed more
light on the non-perturbative dynamics of gauge theories and strings.

\item
Here we considered aspects of this problem in some simple, yet probably
significant cases.

\end{itemize}

\end{enumerate}

\section*{Acknowledgments}
A.G. is on leave of  absence from the Racah Institute of Physics, The
Hebrew University, Jerusalem 91904, Israel; the work of A.G.
is supported in part by BSF - American-Israel Bi-National
Science Foundation, and by the BRF - the Basic Research Foundation.
The work of MP is supported in part by NSF under grant PHY-9318781.


\section*{References}
\begin{thebibliography}{99}

\bibitem{o} D.I. Olive, hep-th/9508089.
\bibitem{s} K. Intriligator and N. Seiberg, hep-th/9509066.
\bibitem{gpr}  A. Giveon, M. Porrati and E. Rabinovici,
              hep-th/9401139, {\em Phys. Rep.} 244 (1994) 77.
\bibitem{sen} A. Sen, hep-th/9402002.
\bibitem{ht} C.M. Hull and P.K. Townsend, hep-th/9410167, \NPB 438 (1995) 109.
\bibitem{p} J. Polchinski, hep-th/9607050.
\bibitem{gmr} A. Giveon, N. Malkin and E. Rabinovici, \PLB 220 (1989) 551.
\bibitem{gp} A. Giveon and M. Porrati, hep-th/9605118.
\bibitem{ggpz}  L. Girardello, A. Giveon, M. Porrati and A. Zaffaroni,
                hep-th/9406128, \PLB 334 (1994) 331;
                hep-th/9502057, \NPB 448 (1995) 127.
\bibitem{dvv}  R. Dijkgraaf, E. Verlinde and H. Verlinde, in
           {\em Perspectives in String Theory, proceedings} (Copenhagen, 1987)
            117.


\end{thebibliography}
\end{document}